\documentclass[reprint,amsmath,amssymb,aps,prb,superscriptaddress]{revtex4-2}
\usepackage{natbib}
\usepackage{graphicx}
\usepackage{dcolumn}
\usepackage{bm}
\usepackage{subfiles}
\usepackage{hyperref}
\usepackage[compact]{titlesec}
\usepackage{float}

\begin{document}

\preprint{APS/123-QED}

\title{Evidence of 3\textit{d}-4\textit{f} antiferromagnetic coupling in strain-tuned PrCo$_{0.5}$Ni$_{0.5}$O$_{3-\delta}$ epitaxial films }

\author{P. K. Sreejith}
\thanks{sreejithpksr@gmail.com}
\affiliation{Department of Physics, Low Temperature Physics Laboratory,  Indian Institute of Technology Madras, Chennai 600036, India}
\affiliation{Department of Physics, Nano Functional Materials Technology Center, Material Science Research Center, Indian Institute of Technology Madras, Chennai 600036, India}
\author{H. B. Vasili}
\affiliation{School of Physics and Astronomy, W. H. Bragg Building, University of Leeds, Leeds LS2 9JT, UK}
\author{W. Li}
\affiliation{ALBA Synchrotron Light Source, E-08290 Cerdanyola del Vallès, Barcelona, Catalonia, Spain}
\author{M. Valvidares}
\affiliation{ALBA Synchrotron Light Source, E-08290 Cerdanyola del Vallès, Barcelona, Catalonia, Spain}
\author{G. Burnell}
\affiliation{School of Physics and Astronomy, W. H. Bragg Building, University of Leeds, Leeds LS2 9JT, UK}
\author{O. Cespedes}
\affiliation{School of Physics and Astronomy, W. H. Bragg Building, University of Leeds, Leeds LS2 9JT, UK}
\author{K. Sethupathi}
\affiliation{Department of Physics, Low Temperature Physics Laboratory,  Indian Institute of Technology Madras, Chennai 600036, India}
\affiliation{Quantum Centre of excellence for Diamond and Emergent Materials (QuCenDiEM), Indian Institute of Technology Madras, Chennai 600036, India}
\author{V. Sankaranarayanan}
\affiliation{Department of Physics, Low Temperature Physics Laboratory,  Indian Institute of Technology Madras, Chennai 600036, India}
\author{M. S. Ramachandra Rao}
\thanks{msrrao@iitm.ac.in}
\affiliation{Department of Physics, Nano Functional Materials Technology Center, Material Science Research Center, Indian Institute of Technology Madras, Chennai 600036, India}
\affiliation{Quantum Centre of excellence for Diamond and Emergent Materials (QuCenDiEM), Indian Institute of Technology Madras, Chennai 600036, India}
\date{\today}

\begin{abstract}
The strong exchange interaction between 3\textit{d}-4\textit{f} magnetic sublattice in rare-earth perovskites introduces a variety of complex magnetic states hosting fascinating electronic ground states with exotic properties. Especially when it comes to rare-earth nickelate and cobaltite perovskites, tuning their rich magnetic phase diagram and spin-state transitions make them potential candidates for spintronic applications. Here, we report the observation of antiferromagnetic coupling between Pr 4\textit{f} and Ni/Co 3\textit{d} magnetic sublattices and its tunability with strain in PrCo$_{0.5}$Ni$_{0.5}$O$_{3-\delta}$ (PCNO) thin films. SQUID magnetization measurements reveal ferromagnetic (FM) ordering around 25 K, followed by a spin glass transition at low temperatures subject to spin reorientation. Competing magnetic interactions arise owing to the 3\textit{d}-4\textit{f} antiferromagnetic (AFM) coupling between Pr and Co/Ni sublattice as revealed by the X-ray absorption spectroscopy (XAS) and X-ray magnetic circular dichroism (XMCD) at the Pr $M_{4,5}$ and Co/Ni $L_{2,3}$ absorption edges. Strain dependence on these AFM coupling reveals an increase (decrease) in the AFM exchange interaction for tensile (compressive) strained films, leading to a net decrease (increase) in the magnetization of PCNO films at low temperatures. The relative increase in low-temperature negative magnetoresistance for compressively strained films also reflects the enhanced ferromagnetic ordering in the system. The angle-dependent magnetoresistance measurements reveal a two-fold anisotropic magnetoresistance (AMR) in tensile strained PCNO films. In contrast, temperature-dependent switching of AMR accompanied by a two- to four-fold symmetry crossover is observed for LaAlO$_3$-grown compressive strained films. This study highlights the tunability of 3\textit{d}-4\textit{f} magnetic exchange interactions in rare-earth cobaltate-nickelate systems, providing deeper insights into the fundamental nature of magnetism in such complex materials.

\end{abstract}

\maketitle

\section{Introduction}
%
Exchange interactions between transition metal and rare-earth element in perovskite system play a prominent role in enhancing the magnetocrystalline anisotropies through the interplay of large spin-orbit coupled localized \textit{f}- and the itinerant \textit{d}-orbitals \cite{McCallum2014}. 
Tailoring the 3\textit{d}-4\textit{f} magnetic interactions could effectively open up many exciting frontiers in engineering magnetic materials with vast potential for spintronic applications. Rare-earth cobaltites and nickelates are potential model systems for studying 3\textit{d}-4\textit{f} exchange interactions, hosting a variety of exotic properties, including metal-insulator and spin-state transitions, superconductivity, non-collinear magnetism and many novel topological phases \cite{medarde1997structural,korotin1996intermediate,tokura1998thermally,ren2011spin,catalano2018rare,li2019superconductivity}. Metal-insulator (MI) transitions and complex antiferromagnetic (AFM) ordering observed in rare-earth nickel perovskites (\textit{R}NiO$_3$) are closely associated with charge disproportionation among the neighbouring Ni-ions \cite{medarde1997structural,catalano2018rare}. MI and AFM ordering temperatures in \textit{R}NiO$_3$ exhibit a strong dependence on the rare-earth (\textit{R}) cation sizes as evident from the nickelate phase diagram \cite{catalano2018rare}. Bulk LaNiO$_3$, having the largest lanthanide cation, remains a paramagnetic metal throughout the temperature range. While, for lighter rare-earths such as PrNiO$_3$, both MI and AFM transitions occur at 130 K \cite{lacorre1991synthesis}. This results from the direct correlation of rare-earth cation size to Ni-O-Ni bond length and bond angle, which in turn determine the magnetic exchange coupling between neighbouring spins. On the other hand, Co-ions in rare-earth cobalt oxides (\textit{R}CoO$_3$) possess a unique ability to switch between different spin states modulated by temperature and strain parameters, owing to the comparable crystal field splitting energy and Hund's exchange energy. The Co-spin states traverse through the non-magnetic low spin (LS) state at low temperatures to the intermediate spin state and then to high spin (HS) states with an increase in temperature\cite{korotin1996intermediate,tokura1998thermally,ren2011spin}. Solid solutions of rare-earth cobaltites and nickelates utilize the spin state transitions in Co-ions and charge disproportionation of Ni-ions to tune the transport and magnetic properties of the combined system, thereby generating unique properties such as magnetic spin glass behaviour in \textit{R}Co$_{1-x}$Ni$_x$O$_3$  and temperature dependent anisotropic magnetoresistance switching in LaCo$_{0.5}$Ni$_{0.5}$O$_{3-\delta}$ thin films \cite{Viswanathan2009,Hammer2004,tomevs2011transport,Sreejith2023}. The rare-earth magnetism also plays an important role in determining the relative orientation of 3\textit{d}-moments, thereby directly affecting the magnetic anisotropy in epitaxial thin films.
Reduced dimensionality of these ultra-thin film systems causes confinement and strain effects, creating novel states in the system. For example, the tensile strained PrNiO$_3$ thin film shows charge ordering and spin density wave phase achieving a high degree of orbital polarization through strain engineering \cite{Hepting2014}. 

Here, we investigate the 3\textit{d}-4\textit{f} exchange coupling in PrCo$_{0.5}$Ni$_{0.5}$O$_{3-\delta}$ (PCNO) thin films employing polarized X-ray absorption spectroscopy, and SQUID magnetometry and explore its evolution with tensile and compressive strain. Further, with field-dependent electrical transport measurements, we investigate the strain-dependent switching in anisotropic magnetoresistance and localization effects in magnetoresistance and electronic transport properties of epitaxially grown PCNO thin films. 
The organization of the manuscript is as follows: Section II describes the experimental methods used for this study; Section III consists of results and discussion, which is divided into five subsection parts; Part A discusses the crystal structure and morphology of thin films; Part B discusses the overall DC SQUID magnetization results of PCNO films with different strain conditions; in part C, we further investigate the 3\textit{d}-4\textit{f} exchange interaction and its dependency on strain with XAS and XMCD measurements; Part D discusses the strain dependence on the electrical transport properties including MR and AMR. Finally, section IV gives the concluding remarks.


%
\section{Experimental methods}

Polycrystalline PCNO target for thin film deposition was synthesized by the conventional citrate-based sol-gel method, which was then cold pressed into a pellet and sintered at 1100 ${\rm^o}$C. 
Thin films of PCNO were grown on TiO$_2$ terminated SrTiO$_3$ (100) (STO) and LaAlO$_3$ (100) (LAO) substrates using pulsed laser deposition method. An excimer laser source (KrF - 248 nm), with a laser fluence of 2 Jcm$^{-2}$ and a pulse repetition rate of 2 Hz was used. 
The base pressure of the chamber was maintained at 5 x $10^{-5}$ mbar and deposition was carried out under the continuous flow of high pure oxygen (99.999 \%), maintaining O$_2$ partial pressure at 0.2 mbar during deposition. 
The substrate temperature was fixed at 700 ${\rm^o}$C. Post deposition, the film samples were kept at 100 mbar O$_2$ pressure while cooling down to ensure proper oxygen stoichiometry of the samples. 
The structural characterization of the thin film samples was done using the high-resolution X-ray diffractometer (HRXRD) (Rigaku Smart Lab II) with Cu K$_\alpha$ source ( $\lambda = 1.54$ \AA).
$2\theta-\omega$ scans around the (200) peak were used for conforming the phase formation, and X-ray reflectivity (XRR) measurements were used to determine the film thickness and roughness. 
From the reciprocal space mapping (RSM) around the pseudocubic (103)$_{pc}$ crystal planes, strain parameters were obtained.

The magnetization measurements on PCNO films were carried out using the SQUID vibrating sample magnetometer (SVSM) (MPMS 3, Quantum Design). In order to determine individual elemental contributions to the total magnetic moment, XAS and XMCD measurements were carried at the BOREAS beamline of the ALBA Synchrotron light source \cite{Barla2016}. The data were collected at $M_{4,5}$ absorption edges for Pr and the $L_{2,3}$ edges for Co and Ni ions. To obtain the XMCD spectra, the difference between the XAS of right ($\sigma^+$) and left ($\sigma^-$) circularly polarized light was measured.  Measurements of the sample were made in in-plane geometry with grazing incidence from the film plane under an applied magnetic field of 6 Tesla along the beam direction. The data were collected in the total electron yield mode.

Temperature and field-dependent electrical transport measurements were carried out with the conventional linear four-probe method using the commercially available physical property measurement system (PPMS) from Quantum Design, equipped with a 9 Tesla superconducting magnet. The four probe contacts were aligned along the [100] or [010] crystal axis, which are equivalent positions because of their four-fold cubic symmetry. 


\section{Results and Discussion}
%
%
\subsection{Structural and morphological properties}
\begin{figure*}[!t]
\includegraphics[width=0.9\linewidth]{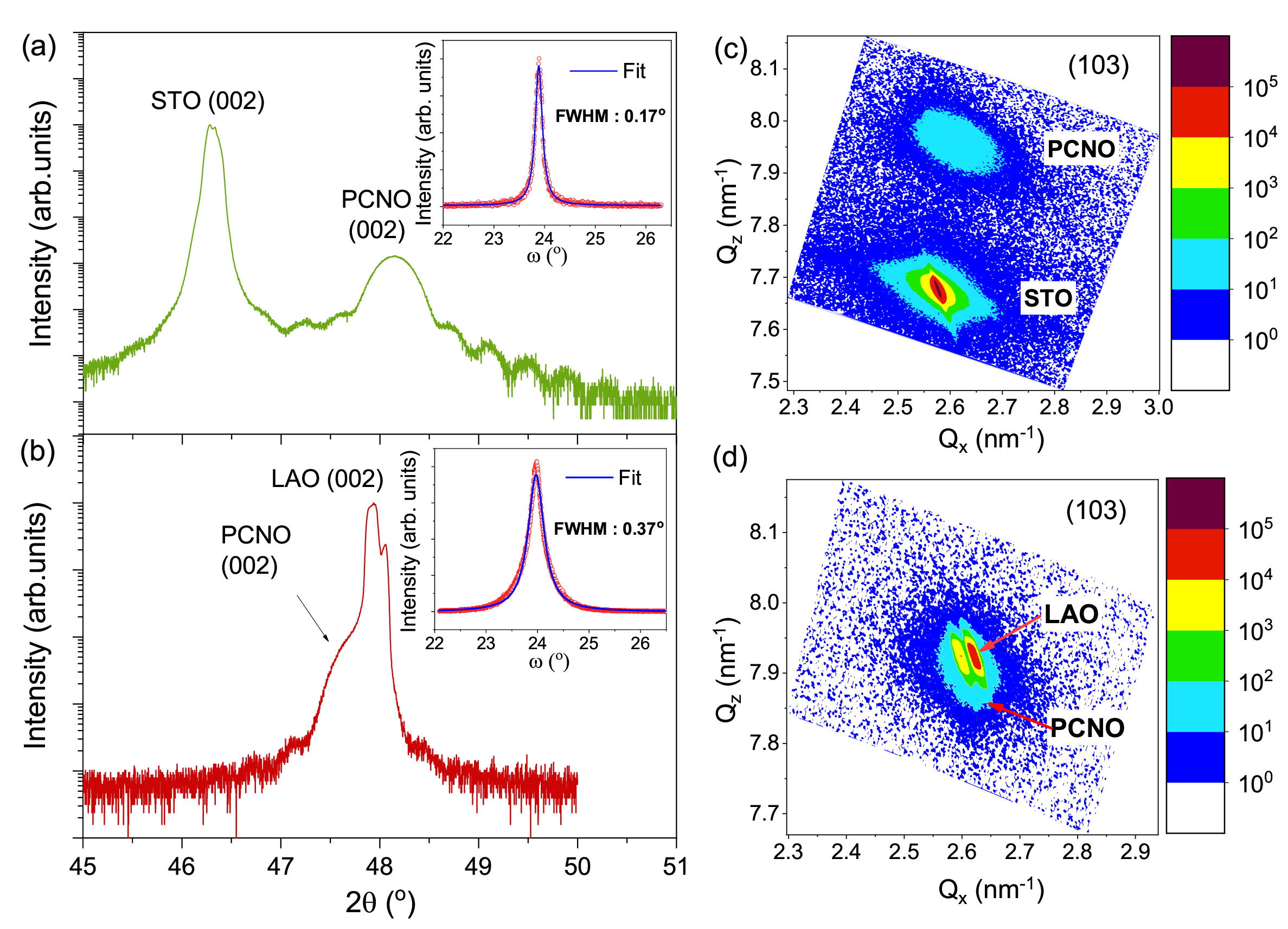}    
\centering
\caption{\label{Fig:xrd} Thin film XRD data showing 2$\theta-\omega$ plots of PCNO thin films around the (002) plane on (a) SrTiO$_3$ and (b) LaAlO$_3$ substrates. Their respective inset shows the corresponding rocking curve graphs. (c) and (d) show the reciprocal space maps along the (103) pseudo cubic plane for PCNO grown on SrTiO$_3$ and LaAlO$_3$ substrates, respectively.}
\end{figure*}
Bulk PCNO has a distorted perovskite structure with an orthorhombic $Pbnm$ phase. Reduced dimensionality and epitaxial strain in PCNO can alter the electronic ground states effectively by changing the $t_{2g}$ and $e_g$ orbital degeneracies. Here, we demonstrate strain tuning of PCNO films under tensile and compressive strain environments. Figures \ref{Fig:xrd} (a) and (b) show the X-ray diffraction scans (2$\theta$-$\omega$) along the [002] direction for PCNO films grown on STO (PCNO//STO) and LAO substrates (PCNO//LAO), respectively. All the films are found to be epitaxial and exhibit good crystalline quality as evidenced by the observation of Kiessig fringe patterns in 2$\theta$-$\omega$ graph as well as the narrow full width at half maximum (FWHM) for the (002) sample peak (FWHM for STO = 0.37$ {\rm^o}$ and for LAO = 0.17$ {\rm^o}$), which is shown in the respective figure insets of (a) and (b). From the XRR fittings (not shown here) we have estimated the film thickness to be around 25 nm for STO and 23 nm for PCNO//LAO. Both the films have an average surface roughness of less than 5 \AA, again confirming the good crystalline quality of the films. From 2$\theta$-$\omega$ scans, it is evident that PCNO//STO film is tensile strained, whereas PCNO on LAO is compressively strained. The observed (002) peak splitting for the LAO substrate is due to the crystal twinning defect, which significantly impacts the magnetocrystalline anisotropy of PCNO films discussed later in this paper. From the 2$\theta$-$\omega$ scans, the c-axis lattice parameter PCNO films are found to be 3.76 \AA\ and 3.81 \AA\ for PCNO//STO and PCNO//LAO, respectively. Compared to this, the pseudocubic lattice parameter for the bulk PCNO is 3.80 \AA.
For a better understanding of the in-plane and out-of-plain strain parameters, RSMs were done along the (103)$_{pc}$ (pc= pseudocubic) Bragg planes. Figures (c) and (d) depict the corresponding RSM images for PCNO//STO and PCNO//LAO films, respectively. From the RSM plots, the observed out-of-plane and in-plane lattice parameters for PCNO//STO film are 3.768 \AA\ and 3.804 \AA\, respectively. Meanwhile, as observed from Fig. \ref{Fig:xrd} (b), the close proximity of the sample peak to that of the substrate makes the RSM for the PCNO//LAO film hardly distinguishable and makes the quantification of the strain parameters difficult. Nevertheless, a qualitative analysis of the RSM shows a slightly compressive nature for the PCNO film.
The RSM analysis shows the strain of PCNO//STO is tensile in nature with an out-of-plane strain of 3.71\%. The film might be partially relaxed along the in-plane direction as evidenced by the slight right shift along Q$_x$, giving an in-plane tensile strain of 1.06\%. 
From the 2$\theta$-$\omega$ plot in Fig. \ref{Fig:xrd} (b), the out-of-plane strain parameter for PCNO//LAO film is determined to be 0.3\%. 
In summary, we can conclude that the PCNO film is tensile strained in STO and compressive strained in LAO, with the out-of-plane lattice parameter of PCNO//LAO film closer to the pseudocubic bulk lattice parameter. In the forthcoming sections below, we will investigate the effect of strain on the electrical and magnetic properties of PCNO films.

\subsection{Magnetic properties}

Magnetic properties of bulk PCNO are greatly influenced by the temperature-dependent cobalt spin state transitions, which in turn depends on the interplay of the crystal field splitting and intra-atomic exchange interactions \cite{Fuchs2008}. In PCNO, the Pr$^{3+}$ being in the 4\textit{f}$^2$ electronic state also contributes to the net magnetization along with Co- and Ni-ions. The magnetic exchange interactions between Pr-4\textit{f} and Ni/Co-3\textit{d} spins will play a pivotal role in the determination of low-temperature magnetic ground states. On the other hand, the inter-atomic Pr $4f-4f$ exchange interactions will be negligibly small owing to the localized nature of 4\textit{f}-orbitals \cite{huang2023hard}. Magnetization measurements on bulk PrCo$_{1-x}$Ni$_x$O$_3$ have revealed that with the increase in Ni doping, the system goes from paramagnetic to spin glass state \cite{tomevs2011transport}. The competing FM and AFM exchange interactions result from the charge transfer from Co to Ni of the form Co$^{3+}$ - Ni$^{3+}$ $\rightarrow$ Co$^{4+}$ - Ni$^{2+}$ giving rise to the glassy phase. 
\begin{figure}[ht]
\includegraphics[width=\linewidth]{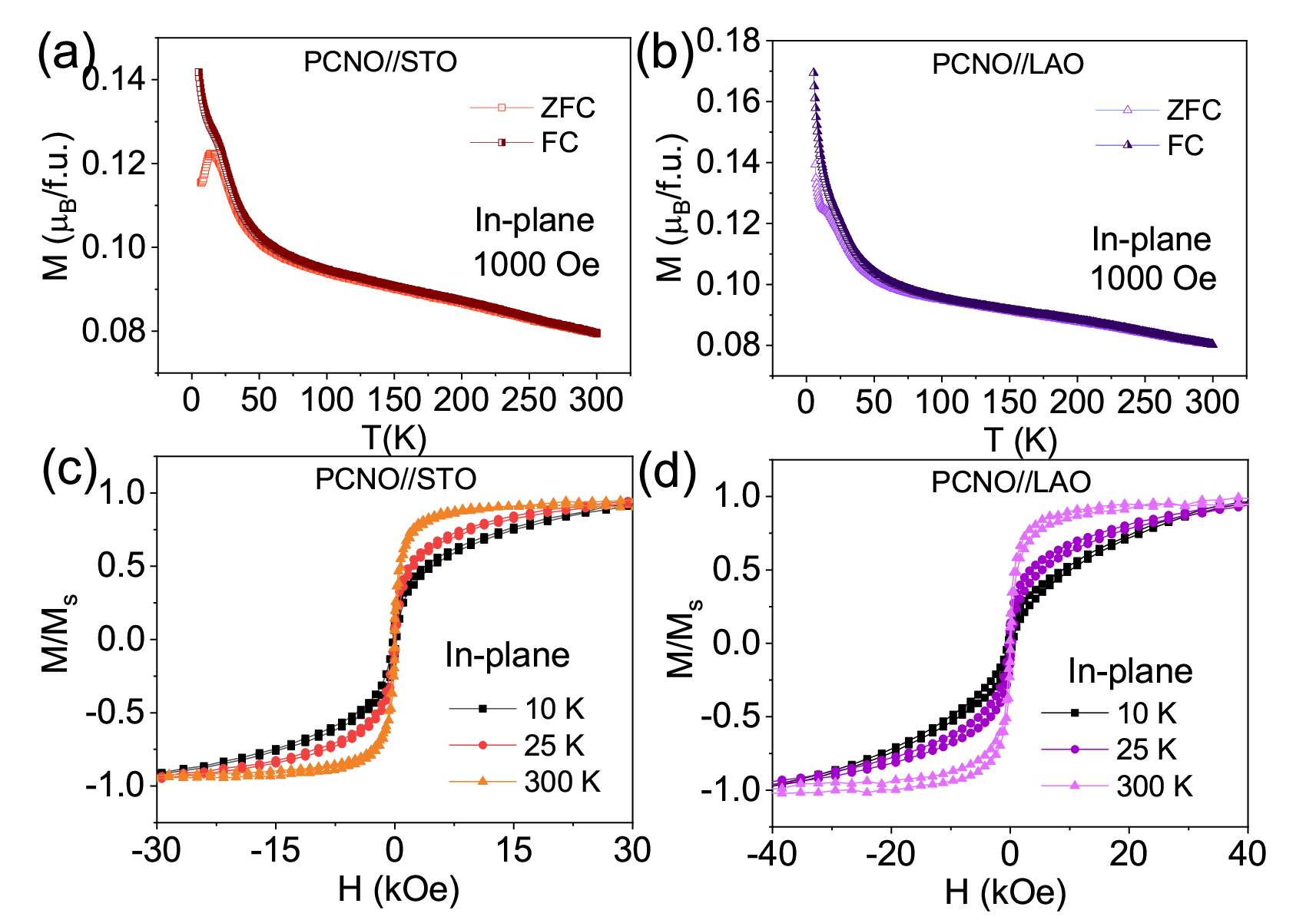}
\caption{\label{fig:MTMH} Temperature-dependent magnetization measurements at zero field-cooled (ZFC) and field-cooled (FC) configuration at an applied field of 1 kOe for PCNO films grown on (a) STO and (b) LAO substrates. The normalized magnetization (M/M$_s$) versus field dependence for PCNO//STO and PCNO//LAO, respectively, are shown in (c) and (d). The diamagnetic contributions from the substrates have been subtracted from the plots}
\end{figure}
For PCNO, the magnetic properties can be significantly altered by low dimensionality and epitaxial strain. Consider, for instance, the case of PrCoO$_3$, which is paramagnetic in bulk form but exhibits ferrimagnetic behavior when subject to tensile strain in thin film form. This change is attributed to the increased presence of HS Co-ions in the ground state under epitaxial strain \cite{Mehta2013}. The change in the HS/LS ratio of Co-ions with strain can be linked to the modification in the crystal field splitting energy ($\Delta_{CF}$) pertaining to its strong dependence on the inter-atomic distances ($\Delta_{CF} = r^{-5}$) \cite{sherman1988}. The effect of tensile and compressive strains on the magnetic ground states in PCNO thin films was studied using temperature (M(T))- and field (M(H))-dependent magnetization measurements in the in-plane geometry (i.e. Magnetic field parallel to the film ab-plane). The substrate diamagnetic contribution to the magnetic moment, obtained from the fitting of the high-field M(H) data, has been subtracted from the sample magnetization data.

Figure \ref{fig:MTMH} (a) and (b) shows the temperature-dependent in-plane magnetization measurements M(T) at 1000 Oe in zero field cooled (ZFC) and field cooled (FC) environments for PCNO//STO (tensile) and PCNO//LAO (compressive) films respectively. M(T) ZFC curve reveals a broad cusp around 14 K followed by a dip at low temperatures in tensile strained PCNO film, showcasing the spin glass type transition similar to bulk PCNO, while for compressive strained film, the ZFC M(T) curve forms a plateau below 17 K, followed by an up turn. From the dM/dT minima, the Curie temperature, $T_\mathrm{C}$ is found to be 26 K and 25 K for PCNO//STO and PCNO//LAO, respectively. The M(T) FC curves in both cases show an upturn at low temperatures where the magnetic contributions from Pr$^{3+}$ ions are prominent. Comparing tensile and compressive strained films, we can clearly observe that the FM long-range ordering is more pronounced in compressively strained films on LAO.   
\begin{figure*}[ht]
    \centering \includegraphics[width=\linewidth,keepaspectratio]{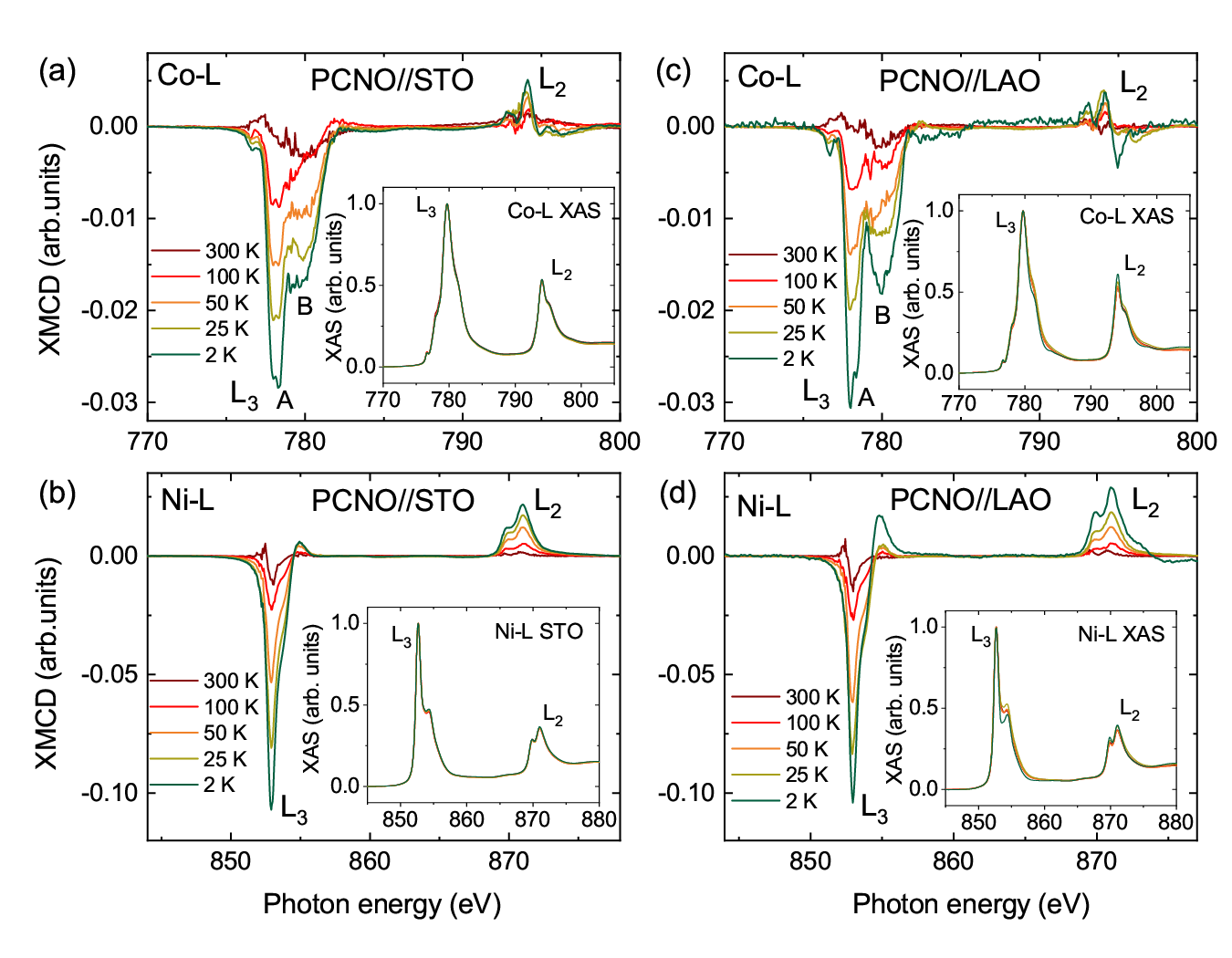}
\caption{\label{Fig:XAS_Co-Ni_L23}(a) and (b) show the XMCD spectra of Co and Ni $L_{2,3}$ edges at various temperatures, respectively, for PCNO//STO film. (c) and (d) shows the respective Co and Ni $L_{2,3}$ edges for PCNO//LAO film at various temperatures. The inset shows their respective XAS spectra at the $L_{2,3}$ edges for various temperatures.}
\end{figure*}
The FM interactions in strained PCNO films are also evident from the in-plane field-dependent M(H) measurements given in appendix Fig. \ref{Fig:MH} as well as in Fig.\ref{fig:MTMH} (c) and (d) for PCNO//STO and PCNO//LAO films, respectively. The presence of FM saturation magnetization at 300 K is a clear indication of the presence of FM clusters at room temperature, well above the magnetic transition temperature. This is similar to the short-range FM domains observed in SmNi$_{1-x}$Co$_x$O$_3$ system in the range 0.3 \textless\ x \textless\ 0.8, inducing a spin-glass state in the system \cite{perez1999electronic}. Figures (c) and (d) show the normalized in-plane isothermal field-dependent magnetization (M/M$_s$) curves at various temperatures for PCNO//STO and PCNO//LAO, respectively. In both the films, the in-plane saturation fields increase as the temperature is lowered, which hints towards a spin-reorientation transition as expected in spin glass systems \cite{Sreejith2023}.

For PCNO films, the saturation magnetization M$_s$ at 10 K is 0.46 $\mu_B$/f.u. for STO-grown film, whereas it is 0.74 $\mu_B$/f.u. for LAO-grown film as seen from Fig. \ref{Fig:MH} in the appendix. This is in agreement with earlier reports showcasing the enhancement of M$_s$ with compressive strain and reduction in M$_s$ with tensile strained films \cite{Zarifi2016,herklotz2013magnetism}. 
The compressive strain introduces distortions in the form of octahedral rotations such that it reduces the Pr and Co/Ni bond lengths, causing the spin to move closer to each other, which leads to enhanced exchange interaction. From the out-of-plane lattice parameters of PCNO//LAO, PCNO//STO and bulk PCNO being 3.81 \AA, 3.76 \AA\ and 3.80 \AA, respectively, it is worth noting that the lattice parameter of the compressively strained PCNO//LAO film is more relaxed towards the bulk PCNO lattice parameter than tensile strained PCNO//STO film, which could also give rise to increased M$_s$, as reported previously \cite{Mehta2011}. The SQUID magnetization measurements hint towards the growth of antiferromagnetic interactions below the transition temperature. In order to get insights into the role of 3\textit{d}-4\textit{f} exchange interactions in the low-temperature magnetic ordering and the strain evolution of spin dynamics of the individual moments in PCNO films, we have performed detailed X-ray absorption spectroscopic studies which are discussed below.

\subsection{X-ray absorption studies}
\subsubsection{\textbf{XAS and XMCD spectroscopy}}

Soft X-ray absorption spectroscopy is a powerful tool for understanding the element- and site-specific electronic and spin transitions in complex oxide systems. Delving into the underlying mechanisms of strain-induced effects on magnetism, especially on the cobalt spin state transitions and Pr-4\textit{f}-Co/Ni-3\textit{d} magnetic exchange interactions in the PCNO system, we have probed the XAS and XMCD at Co-, Ni-$L_{2,3}$ edges and Pr-$M_{4,5}$ edges as a function of temperature. 

Figures \ref{Fig:XAS_Co-Ni_L23} (a,b) and (c,d) show the temperature-dependent XMCD spectra of Co-$L_{2,3}$ and Ni-$L_{2,3}$ edges for tensile and compressive strained films, respectively. We have plotted the XAS data by taking the average of left and right circular polarizations. The insets of the figure show their respective normalized XAS spectra.  The XAS features at either PCNO//STO or PCNO//LAO samples appear to have similar multivalent states throughout the temperature range, 2-300 K, except for some little differences in the intensities for the LAO data, implying largely invariant valent states in the samples. In the Co-$L_3$ XAS data, the fine structure arises from the linear combination of the multiplet structures showing spectral signatures corresponding to Co$^{3+}$ (main peak) and Co$^{4+}$ (left and right shoulder peaks) final states \cite{lin2010local,guillou2017valence}. The small peak at 777 eV corresponds to the Co$^{2+}$ state, arising due to the small number of oxygen vacancies present in the system, mostly at the surface \cite{chen2014complete}. Similarly, at the Ni-$L_3$ edges, the main peak has a prominent shoulder peak at the right-hand side along with the peak doublet at the Ni $L_2$ edge, indicating the mixed valence state of Ni$^{2+}$ and Ni$^{3+}$ ions \cite{piamonteze2015interfacial}. The presence of Co$^{4+}$ and Ni$^{2+}$ thus signifies the charge transfer from Co-3\textit{d} to Ni-3\textit{d} orbitals, causing the ferromagnetic ordering in the system. A detailed analysis of the temperature evolution of the XAS and XMCD line shapes of the Co and Ni-$L$ edges can be found in our earlier work on LaCo$_{0.5}$Ni$_{0.5}$O$_{3-\delta}$ \cite{Sreejith2023}. 

Temperature-dependent XMCD data, on the other hand, give information about the element-specific magnetic interactions as their dichroism reflects the spin polarization of Co, Ni and Pr ions in the system. The XMCD for Co and Ni $L_{2,3}$ edges show a systematic increase in their spectral intensities as samples cool down from 300 K to 2 K. A clear evolution of the valence states can be found for the Co-$L_3$ edge as the peak splits into a doublet marked as A and B at low temperatures, as shown in Fig. \ref{Fig:XAS_Co-Ni_L23} (a,c). Here, the peaks A and B correspond to dominant Co$^{3+}$ and Co$^{4+}$ valence states, respectively, paving the way for double exchange FM interactions in the system. The change in valence states is directly linked to the temperature evolution of the Co spin states as previously reported \cite{Sreejith2023}. On the other hand, the line shape of the Ni $L_{2,3}$ edges is invariant throughout the temperature range, indicating no change in the ratio of Ni$^{3+}$/Ni$^{2+}$ ions with the temperature. The sign of the main XMCD peaks of both Co- and Ni-$L_{2,3}$ edges is negative, which represents that the spin orientation of Co and Ni-magnetic sublattices are parallel and is FM in nature. However, the magnetic features of the Pr ions are radically different and are discussed in the following section. 

\begin{figure*}[ht]
    \centering \includegraphics[width=\linewidth,keepaspectratio]{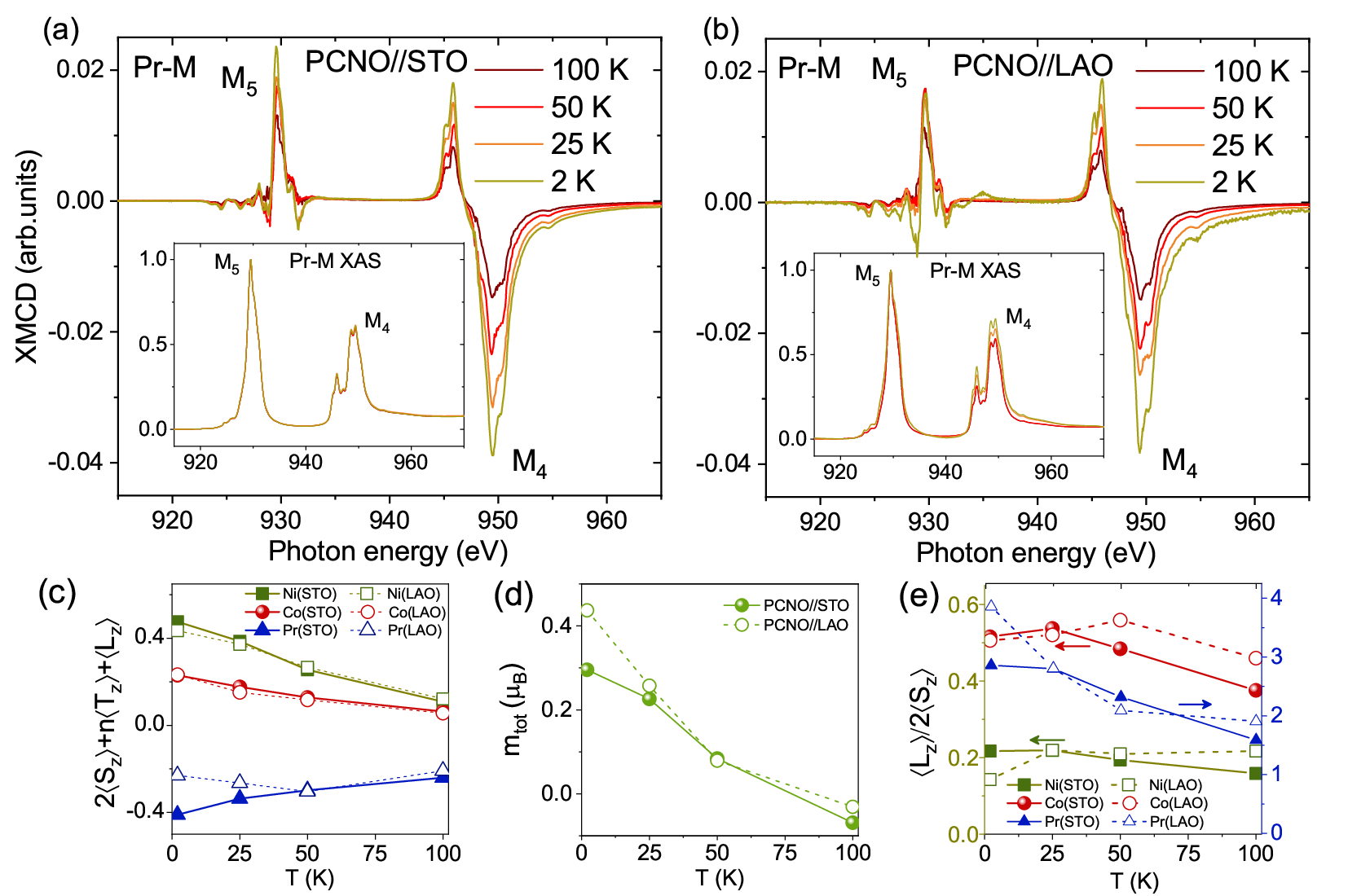}
    \caption{\label{Fig:XAS_Pr_L23}XMCD spectra of Pr-$M_{4,5}$ edges at different temperatures for (a) PCNO//STO and (b) PCNO//LAO films. The corresponding XAS spectra of Pr-$M_{4,5}$ edges at various temperatures are shown in the respective insets. (c) Temperature dependence of the net magnetic moment, 2$\langle S_z \rangle$ + $n\langle T_z \rangle$ + $\langle L_z \rangle$, for Pr, Ni and Co-ions derived from XMCD sum rules. The solid lines connecting the filled symbols correspond to elemental total moment values for PCNO//STO film, and dashed lines connecting open symbols correspond to moment values of PCNO//LAO film. (d) Element specific ratio of orbital to spin angular moment, $m_l/m_s$ = $\langle L_z \rangle$/2$\langle S_z \rangle$, for strained PCNO films. The respective substrates are shown in the parenthesis. (e)  Temperature dependence of total magnetic moment calculated from XMCD spectra for PCNO//STO (filled spheres connected by solid lines) and PCNO//LAO (open circles connected by dashed lines)}
\end{figure*}

\subsubsection{\textbf{The curious case of Pr magnetism}}
At low temperatures, the rare-earth Pr$^{3+}$ ($f^2$) electronic spins order magnetically under the influence of the strong molecular field due to 3\textit{d}-3\textit{d} magnetic exchange interactions. Unlike 3\textit{d}-transition metals, the 4\textit{f}-elements show strong orbital moments because of their localized nature. Figure \ref{Fig:XAS_Pr_L23} insets in (a) and (b) shows the XAS profiles of Pr-$M_{4,5}$ edges for tensile and compressive strained PCNO films, respectively. The XAS line shape of Pr-$M_{4,5}$ edges shows a rich multiplet structure corresponding to the Pr$^{3+}$ valence state due to the strong Coulomb interaction and hybridization with oxygen 2\textit{p}-states \cite{Bernal-Salamanca2022}. The Pr-$M_{4,5}$ XAS data in Fig. \ref{Fig:XAS_Pr_L23} (a,b) is similar to those of the parent compounds, PrCoO$_3$ \cite{Padilla2014} and PrNiO$_3$ \cite{Thole1985} systems, confirming a predominant Pr$^{3+}$ valency.

On the other hand, the XMCD spectra for Pr-$M_{4,5}$ edges (Fig. \ref{Fig:XAS_Pr_L23} (a,b)) exhibit opposite sign compared to those observed at the Co/Ni $L_{2,3}$ edges, hinting towards an anti-parallel alignment of spins in the Pr sub-lattice to that of Co/Ni sublattice. This is contradictory to the general notion of Pr sublattice aligning with the magnetic field at low temperatures, as reported in bulk PCNO \cite{tomevs2011transport}. However, from the XMCD spectra, it is also evident that the orbital moment is much larger and opposite to the spin moment. The large orbital moment is expected for \textit{f}-orbitals, as they are localized and protected from the crystal field effects, and there will be no quenching of orbital moments. Pr$^{3+}$ has a 4$f^2$ configuration, which is less than half-filled; hence, the spin and orbital moments are opposite to each other. It is very important to note that the net magnetic moment of the Pr sublattice also has a strong intra-atomic magnetic dipole contribution along with spin and orbital moments. We have determined the spin, orbital, and magnetic dipole moments of all the constituting ions using the XMCD sum rules; the discussion follows in the next section.
 
\subsubsection{\textbf{XMCD sum rules}}

XMCD sum rules are developed by Thole and Carra for calculating the individual magnetic moments of the ground state \cite{Thole1985,Thole1992,Carra1993,Chen1995XMCD,Vasili2017}. These calculations are based on the electric dipole transitions from the spin-orbit-split core level to the valence shell electrons carrying the shell-resolved operators. For the 3\textit{d}-transition element edges such as Co/Ni $L_{2,3}$ ($2p\rightarrow3d$) absorption edges, these sum rules are represented as
\begin{eqnarray}
\label{eqn:sum_rule_L}
2\langle S_z \rangle + 7\langle T_z \rangle = \left(\frac{6p-4q}{r}\right)N_h, \langle L_z \rangle = \left(\frac{4q}{3r}\right)N_h\nonumber\\
\end{eqnarray}
Similarly, for the rare-earth, Pr-$M_{4,5}$ ($3d\rightarrow4f$) absorption edge is given as
\begin{eqnarray}
    \label{eqn:sum_rule_M}
    2\langle S_z \rangle + 6\langle T_z \rangle = \left(\frac{5p-3q}{r}\right)N_h, \langle L_z \rangle = \left(\frac{2q}{r}\right)N_h\nonumber\\
\end{eqnarray}
where $\langle S_z \rangle$, $\langle L_z \rangle$ and $\langle T_z \rangle$ are the expectation values of the \textit{z} component of the spin angular moment, orbital angular moment and the magnetic dipole operator, respectively, of the 3\textit{d} and 4\textit{f} electrons \cite{Vasili2017}. The parameter \textit{r} represents the integrated value of the total XAS at the $L_2$ and $M_4$ post edges (post subtraction of the step function), and \textit{p} and \textit{q} parameters are the integrated values of the XMCD at $L_3(M_5)$ and $L_2(M_4)$ post edges respectively. $N_h$ is the number of holes for the 3\textit{d} and 4\textit{f} shells of the respective Co, Ni and Pr ions.
From equation \ref{eqn:sum_rule_L} and \ref{eqn:sum_rule_M}, we can calculate the spin and orbital magnetic moments  as $m_s = -2\langle S_z \rangle$ $\mu_B$ and $m_l = -\langle L_z \rangle$ $\mu_B$, respectively. 
The total net magnetic moment of each ion is (2$\langle S_z \rangle$ + n$\langle T_z \rangle$ + $\langle L_z \rangle$). Here, the coefficient of $\langle T_z \rangle$, n = 7 for 3\textit{d} transition elements and n = 6 for the rare-earth elements. The following comments are relevant to the sum rule data analysis. Corrections to the spin sum rules were taken into account for the Ni and Co ions, as discussed in Ref. \cite{Piamonteze2009}. Additionally, the values of $\langle T_z \rangle$ for Co/Ni ions are assumed to be negligible due to the cubic symmetry of the \textit{d} orbitals. However, the $\langle T_z \rangle$ cannot be negligible for the rare-earth ions (in this case, the Pr$^{3+}$ ions) because of the localized and aspherical 4\textit{f} electron charge densities. The magnitude of $\langle T_z \rangle$ cannot be determined directly; therefore, we estimate the spin-only component of the Pr magnetic moment by assuming that the $\langle T_z \rangle$ is proportional to $\langle S_z \rangle$. The ratio of $\langle T_z \rangle$/$\langle S_z \rangle$ = 8/5 was taken for Pr$^{3+}$ using the theoretical expectation values for $\langle T_z \rangle$ and $\langle S_z \rangle$ from the expectation values $\langle w^{xyz} \rangle$ given in Ref. \cite{van_der_Laan1996}. Using the above ratio, we have extracted the spin-only magnetic moment 2$\langle S_z \rangle$ of the Pr ions.

Figure \ref{Fig:XAS_Pr_L23} (c) shows the net moment contributions of the individual Ni, Co and Pr ions as a function of temperature. Data show that the net Pr moment is negative with respect to the Co- and Ni-ions. This confirms the AFM coupling of the Pr moments to the Ni/Co moments. It is also evident that the Ni moments are larger than that of the Co moments for any given temperature. The total magnetic moment, $m_{tot}$ = $m_\mathrm{Co}$ + $m_\mathrm{Ni}$ + $m_\mathrm{Pr}$, is plotted in Figure \ref{Fig:XAS_Pr_L23} (d) for the PCNO films on both STO and LAO substrates. This data reveals remarkable changes below 50 K, supporting the strain-induced spin-reorientation transition below 25 K. The discussion is as follows.

\subsubsection{\textbf{Strain effects: XMCD analysis}}

The XMCD spectral features and the derived magnetic moments have a clear influence from the substrate-induced strain in the PCNO films. Let us first discuss the XMCD spectra. The peak B labelled in the Co-$L_3$ XMCD data shown in Fig. \ref{Fig:XAS_Co-Ni_L23} (a,c), which corresponds to the Co$^{4+}$ state, appears more prominent in the compressively strained films compared to the tensile ones. The PCNO//STO data exhibit a maximized ratio of Co$^{4+}$/Co$^{3+}$ at 25 K and decreases at the base temperature. This reflects the cusp seen in the M(T) ZFC curve of PCNO//STO in the SQUID magnetometry data. On the other hand, the PCNO//LAO data show a monotonic increment at low temperatures, as seen in the SQUID data. 
Though the net magnetic moments of Co and Ni ions, shown in Figure \ref{Fig:XAS_Pr_L23} (c), appear similar and monotonously increasing at lower temperatures for both LAO and STO substrates, the spin reorientation transition below 25 K is clearly visible from the lowering of $\langle L_z \rangle$/2$\langle S_z \rangle$ values for Co and Ni ions at 2 K, irrespective of the substrates (see Figure \ref{Fig:XAS_Co-Ni_L23} (e)). With an applied magnetic field of 6 Tesla, the spin re-orientations are hard to observe in the spin-only moments of Ni and Co as their orbital moments get quenched. However, the orbital-to-spin moment ratio can be sensitive to the spin-reorientation transition.

On the other hand, the Pr-\textit{M} edge for STO- and LAO-grown samples show significant differences in the net magnetic moments and the $\langle L_z \rangle$/2$\langle S_z \rangle$ values, which explains the increased FM double exchange interactions and thus the observed higher saturation magnetization in the compressively strained sample.

The main difference in the Pr data is that the XMCD peak at $L_3$ edge is larger for PCNO//STO than that of PCNO//LAO, resulting in enhanced spin and dipole magnetic moments. We can see this in the derived net Pr moments in Figure \ref{Fig:XAS_Co-Ni_L23} (c), where the Pr moment of the STO sample increases significantly at low temperatures while the LAO sample moment turns down to lower values from 50 K to 2 K in accordance with the spin-reorientation discussed in the SQUID magnetometry data. From Figure \ref{Fig:XAS_Co-Ni_L23} (e), we can see that the $\langle L_z \rangle$/2$\langle S_z \rangle$ values for the Pr ions are one order higher than that of Co and Ni ions. This is from the fact that the localized 4\textit{f} moments exhibit larger orbital moments, whereas the 3\textit{d} moments are nearly quenching. The $\langle L_z \rangle$/2$\langle S_z \rangle$ values of Pr show an enhanced value at 2 K for the compressively strained film, whereas the tensile strained film doesn’t change much. Our data show that Pr magnetism is the main driving force behind the observed strain-dependent magnetization in the PCNO films. The total magnetic moment (by adding Ni, Co, and Pr moments) given in Fig. \ref{Fig:XAS_Pr_L23} (d) reflects the overall magnetization measured by the SQUID magnetometry.  

At low temperatures, the Co spin reorients and is antiferromagnetically coupled to the Pr sublattice. For PCNO//STO, this coupling is enhanced, resulting in the net reduction of magnetic moment. Whereas, for the PCNO//LAO, the 3\textit{d}-4\textit{f} AFM coupling is reduced, resulting in a higher magnetic moment. Reducing the Pr moments at low temperatures for the compressively strained film alters the above-mentioned AFM coupling in the system that fingerprints the observed spin-reorientation transitions below 25 K.

In the next section, we extend our studies to the strain effects on the electrical resistivity and magnetotransport anisotropic properties of PCNO in light of the 3\textit{d}-4\textit{f} magnetic exchange interaction. 

\subsection{Electrical transport properties}

In bulk PrCo$_{1-x}$Ni$_x$O$_3$, the transport mechanism reportedly follows the ES-VRH mechanism below 160 K, showing strong electron localization effects \cite{tomevs2011transport}. Temperature-dependent resistivity measurements on PCNO thin films reveal a crossover from 3D Mott VRH to ES VRH conduction in tensile strained film, while resistivity data of compressive PCNO//LAO film fits perfectly with Mott VRH at low temperatures, followed by high-temperature Arrhenius fit. A detailed analysis of the temperature dependence of resistivity is given in appendix-\ref{appendix}.

\subsubsection{\textbf{Magnetoresistance}}

\begin{figure}[ht]
    \centering
    \includegraphics[width=\linewidth,keepaspectratio]{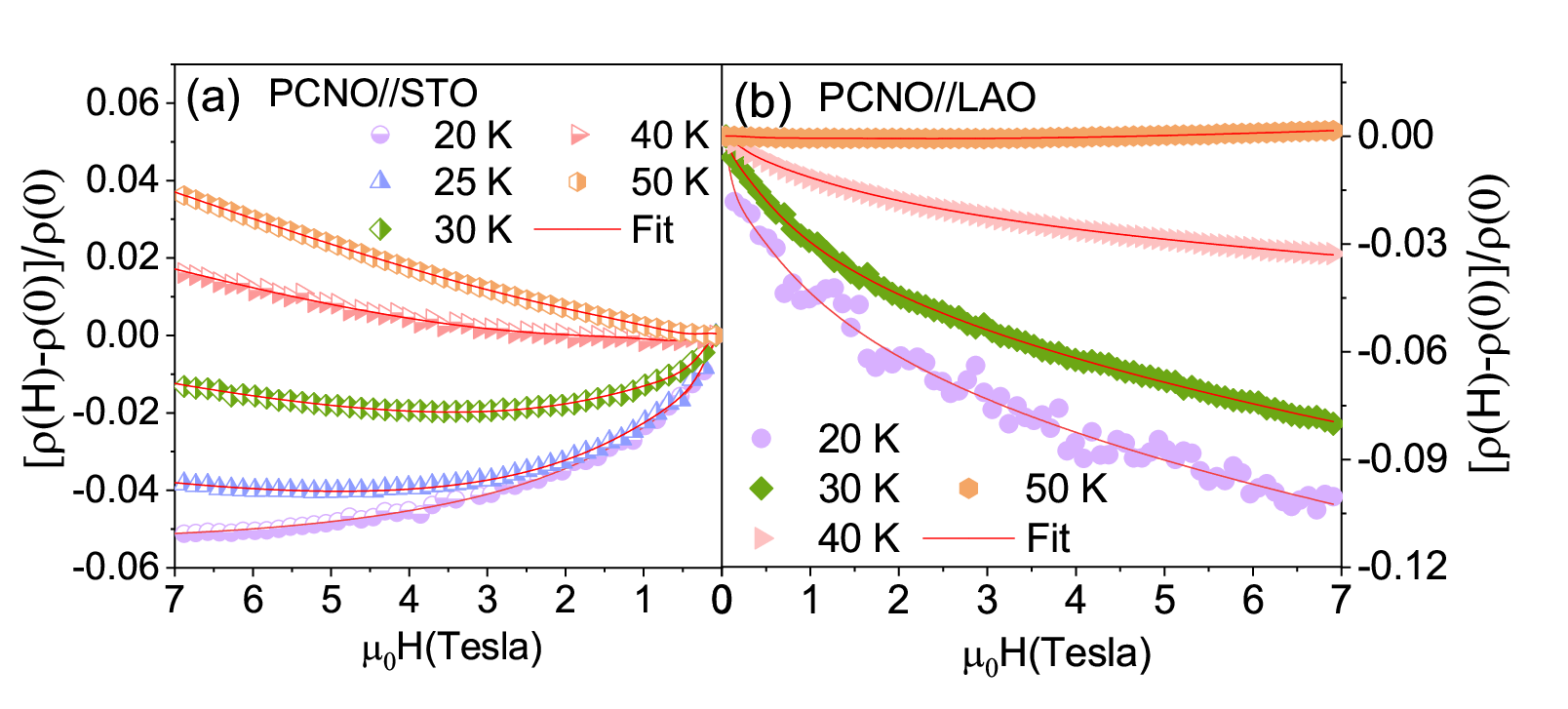}
\caption{\label{Fig:MR} Magnetoresistance $([\rho(H)-\rho(0)]/\rho(0))$ plots of (a) PCNO//STO and (b) PCNO//LAO films and their respective fits using the modified Koshla-Fisher model (\ref{eqn:KF}) at various temperatures. Symbols represent the experimental data, and solid red lines correspond to the fit.}
\end{figure}

Field-dependent resistivity measurements were carried out in strained PCNO thin films. The effect of strain on magnetoresistance (MR) in PCNO films is explored in PCNO//STO and PCNO//LAO films. Figure \ref{Fig:MR} (a) and (b) depict the isothermal magnetoresistance (MR) measurements with applied fields ranging from 0 to 7 Tesla for PCNO//STO and PCNO//LAO films, respectively. All the films show negative to positive crossover of magnetoresistance (MR) at high temperatures and high magnetic fields. The MR results can be mapped to the magnetization measurements, as the negative MR in ferromagnetic material arises due to the reduced scattering of charge carriers by the localized spins in response to their parallel alignment towards the applied external magnetic field \cite{Peters2010}. As a direct consequence, in PCNO//LAO film, the negative MR is stabilized below 50 K, while for PCNO//STO film, we can see an upturn in the negative MR profile at high fields starting from 25 K, which completely transforms to positive MR at 50 K. This could be a consequence of the increased 3\textit{d}-4\textit{f} AFM exchange interactions in tensile strained films, causing a net reduction in ferromagnetic ordering. In comparison to a similar system, LaCo$_{0.5}$Ni$_{0.5}$O$_{3-\delta}$ where 4\textit{f} magnetism is absent, the MR is negative throughout the temperature range \cite{Sreejith2023}. The size of the rare-earth cation also plays a role in this matter. In PCNO, due to the relatively smaller size of the Pr ion, the bond angles and bond distances of the constituent elements get distorted, thereby possibly reducing the net electron localization \cite{Alonso2000}. With an increase in temperature, electrons gain thermal energy and overcome the localization, giving rise to positive MR. Positive MR has a H$^2$ field dependence resulting from the Lorentz force due to the orbital motion of itinerant electrons. 

The MR data were fit to a modified semiempirical Koshla-Fisher (KF) model given by equation \ref{eqn:KF}, which takes into account the scattering of electrons by the localized magnetic moments in the system by incorporating higher-order terms in the perturbation expansion of the exchange Hamiltonian \cite{Khosla-Fischer1970}. The first term in equation \ref{eqn:KF} describes the spin scattered negative MR part, while the second term is added to the original KF model to explain the positive MR part, which depends on the conductivity and mobility of electrons, based on the two-spin channel sub-band model. In this model, the \textit{p}-\textit{d} hybridization induces spin polarization and creates two distinct spin-polarized bands of different carrier mobilities \cite{Watts2000,pippard1989magnetoresistance,Peters2010}. A third term is introduced in the modified KF equation to describe the positive linear dependence of MR in PCNO films. The origin of the linear term can be attributed to the magnetic inhomogeneities present in the system \cite{niu2021large}. Thus, the modified KF equation can be represented as (\ref{eqn:KF})
\begin{eqnarray}
\label{eqn:KF}
\centering\Delta\rho/\rho= -a^2ln(1+b^2H^2) + \frac{c^2H^2}{1+d^2H^2} + nH
\end{eqnarray}
The coefficients in the first term, \textit{a} and \textit{b} in equation \ref{eqn:KF} given by the relation 
\begin{eqnarray}
a^2= A_1JD(\epsilon_F)[S(S+1)+\langle M^2 \rangle]
\end{eqnarray}
\begin{eqnarray}
b^2= \left[1+4S^2\pi^2\left(\frac{2JD(\epsilon_F)}{g}\right)^4\right]\left(\frac{g\mu_B}{\alpha k_BT}\right)^2
\end{eqnarray}
 
\noindent here $A_1$ relates to the ratio between magnetic ($\sigma_J$) to non-magnetic ($\sigma_0$) scattering cross-section through the relation $A_1$ = A$N_A(\sigma_J/\sigma_0)^2$ with A being a proportionality constant, $N_A$ is the Avogadro's number. $J$ is the exchange interaction energy, $D(\epsilon_F$) is the density of states at the Fermi level ($\epsilon_F$), \textit{g} is the Lande factor, \textit{S} is the localized spin magnetic moment, $\langle M \rangle$ is the average magnetization, $k_B$ is the Boltzmann constant and $\alpha$ is a numerical constant which is close to 1.

Similarly, coefficients \textit{c} and \textit{d}, related to the electron mobilities $(\sigma_1,2)$ and conductivities $(\mu_1,2)$ are given as 
\begin{eqnarray}
\label{eqn:c}
c^2=\frac{\sigma_1\sigma_2(\mu_1+\mu_2)^2}{(\sigma_1+\sigma_2)^2}
\end{eqnarray}
\label{eqn:d}
\begin{eqnarray}
d^2=\frac{(\sigma_1\mu_2-\sigma_2\mu_1)^2}{(\sigma_1+\sigma_2)^2}
\end{eqnarray}
and finally, \textit{n} is a proportionality constant from the linear term

The MR data at temperatures 20 K to 50 K are fitted with the modified KF equation, and the respective $a,b,c,d$ and \textit{n} values are given in Table \ref{table:KF}.
\begin{table}[t]
    \caption{MR fit parameters to the modified Koshla-Fisher empirical model for PCNO samples grown on LAO and STO substrates}
    \label{table:KF}
    \centering
    \begin{ruledtabular}
    \begin{tabular}{ccccccc}
    Substrate&T (K)&\textit{a}&\textit{b}&\textit{c}&\textit{d}&\textit{n} \\ \hline\\
    STO&20&0.112&5.324&0.254&1.711&0.0026 \\
    &25&0.128&4.553&0.269&1.546&0.0065 \\
    &30&0.118&3.650&0.210&1.280&0.0074 \\
    &40&0.106&2.611&0.175&1.086&0.0081 \\
    &50&0.079&3.099&0.145&1.206&0.0087 \\
    \hline \\
    LAO&20&0.109&14.301&0.564&3.599&-0.0023\\
    &30&0.095&10.515&0.572&4.467&-0.0027\\
    &40&0.072&4.949&0.164&1.877&-0.000\\
    &50&0.050&2.511&0.079&1.069&0.0015 \\
   
    \end{tabular}
    \end{ruledtabular}
    \label{tab:table1}
\end{table}
If we assume $D(\epsilon_F$) to be constant for the temperature range under consideration, the parameter \textit{a} greatly corresponds to the ratio of magnetic to non-magnetic spin scattering in the system as well as to the magnetic exchange interaction and net magnetization of the system. Table \ref{tab:table1} shows higher values of \textit{a} for PCNO//STO films at high temperatures, which starts reducing below 25 K. Whereas, for PCNO//LAO, it monotonously increases towards low temperatures. The increase in \textit{a} parameter value indicates a stronger magnetic spin-dependent scattering in the system. The \textit{b} parameter, on the other hand, has a strong dependence on the magnetic exchange interaction. The larger values of \textit{b}, especially at 20 K in PCNO//LAO, imply strong ferromagnetic exchange coupling compared to PCNO//STO. The behavior of \textit{a} and \textit{b} parameters at low temperatures, in conjunction with Pr magnetism explored via XMCD analysis, reflects the increased (decreased) antiferromagnetic 4\textit{f}-3\textit{d} interactions in PCNO//STO (PCNO//LAO) film.

In the next section, we further explore the effects of strain on the magnetoresistance anisotropy of PCNO films using angle-dependent magnetoresistance measurements.

\subsubsection{\textbf{Anisotropic magnetoresistance}}

AMR depicts the effect of longitudinal resistivity on magnetization and current direction. Angle-dependent magnetization measurements for PCNO samples were done with the applied current along the y-axis and the applied magnetic field along the yz plane. The percentage AMR is calculated as
\begin{eqnarray}
\label{eqn:AMR1.1}
\frac{\rho(\theta)-\rho(\perp)}{\rho(\perp)}\times 100 \%
\end{eqnarray}
where $\rho(\perp)$ is the sample resistivity when the magnetic field is applied perpendicular to the crystal plane (i.e. along the [001] direction) and $\rho(\theta)$ is the resistivity of the sample when the applied field is at an angle $\theta$. Experimentally obtained AMR can be fitted with a phenomenological equation given as \cite{Li2010a}
\begin{eqnarray}
\label{eqn:AMR3}
\rho_\text{AMR} = A_0 + A_2\cos[{2(\theta + \theta_0)}] + A_4\cos[{4(\theta + \theta_0)}]
\end{eqnarray}
where $A_2$ and $A_4$ are the respective amplitudes of two-fold and four-fold symmetry components and $\theta_0$ accounts for the misalignment due to the sample rotator.
\begin{figure}[ht]
    \centering
    \includegraphics[width=\linewidth,keepaspectratio]{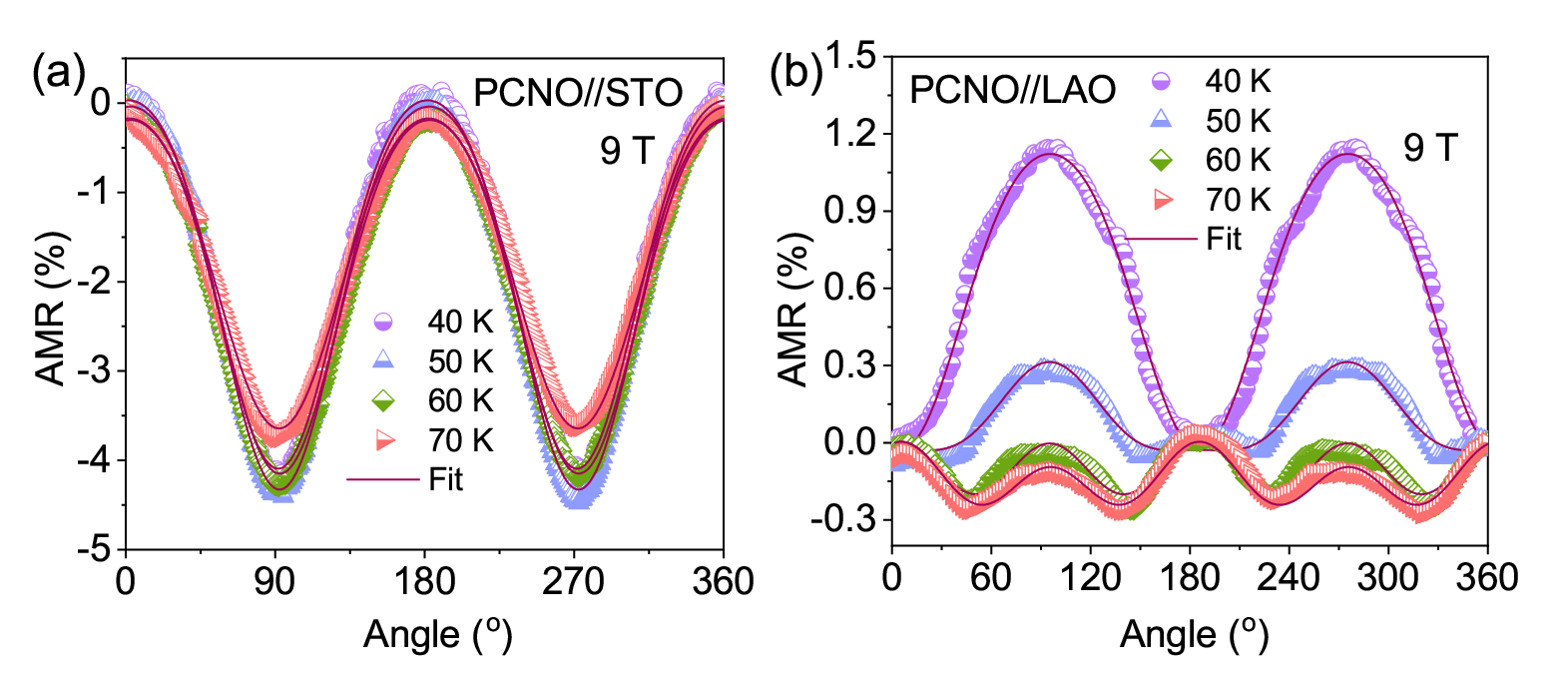}
\caption{\label{Fig:AMR} Anisotropic magnetoresistance measured at various temperatures at an applied field of 9 Tesla for (a) PCNO//STO and (b) PCNO//LAO films.}
\end{figure}

Similar to the doped manganite systems, AMR in PCNO is highly sensitive to the epitaxial strain, which is evident from Fig. \ref{Fig:AMR} \cite{alagoz2015mechanism,wang2010pseudomorphic}. The tensile strained PCNO//STO gives negative AMR, which maximizes when the field is parallel to the film plane, implying the magnetization easy axis lying close to the film in-plane direction, whereas, for compressively strained PCNO//LAO films, especially at low-temperatures, AMR is positive and hence the easy axis is aligned towards the out-of-plane direction. Strain can induce tetragonal distortion on thin films, which lifts the degeneracy of the $e_g$ states. For tensile strained PCNO//STO film, the degeneracy is lifted such as the $d_z^2$ orbital lies higher than the $d_{x^2-y^2}$ orbital and for compressively strained film, PCNO//LAO $d_{x^2-y^2}$ orbital level is higher than $d_z^2$, Hence, promoting preferential orbital occupancy in the $e_g$ levels and influence the magnetic anisotropy of the system \cite{wu2000strain}. Additionally, the large magnetocrystalline anisotropy of the Pr sublattice influences the relative orientation of magnetization easy axis with strain \cite{Kovnir2013}.  

The origin of AMR is generally explained using the two-current model based on the \textit{s}-\textit{d} scattering mechanism, where \textit{s} corresponds to the conduction electrons and \textit{d} to localized electrons. Similar to half metals like LSMO and LCMO, the dominant \textit{s}-\textit{d} scattering mechanism constitutes spin-up conduction electrons to spin-up localized \textit{d}-electrons. In LAO, the crystal twinning could lead to a tetragonal distortion along the [001] direction, which reflects as a dominant four-fold symmetry A$_4$ component in AMR \cite{wang1990twinning}.
Similar to LaCo$_{0.5}$Ni$_{0.5}$O$_3$ films, PCNO also shows strong AMR behaviour, which is highly dependent on strain parameters. In Fig. \ref{Fig:AMR}, temperature-dependent AMR is measured for PCNO films with different strain conditions. It is clearly seen that the sign of AMR is negative for tensile strained film, PCNO//STO. In contrast, for compressively strained PCNO//LAO film, AMR switches signs from positive to negative for temperatures above 50 K. Moreover, in tensile strained film, the magnetocrystalline anisotropy is dominated by two-fold symmetry, while for compressive PCNO//LAO film, there is a temperature-dependent crossover from two-fold to four-fold symmetric AMR. The sign change in AMR can be understood as a consequence of strain-induced spin polarization reversal and changes in the electronic structure\cite{alagoz2015mechanism,chen2009crossover}. The origin of two-fold to four-fold crossover, on the other hand, can be associated with the formation of antiphase boundaries in the twinned LAO substrate. In general, the two-fold and four-fold terms are associated with the cubic and tetragonal crystal field symmetries, respectively. Tetragonal distortion is induced by crystal symmetry \cite{kokado2015twofold,burema2021temperature}.


%
%
\section{Conclusion}

In summary, our study demonstrates the 3\textit{d}-4\textit{f} antiferromagnetic coupling in PCNO thin films, driving the overall magnetization dynamics at low temperatures. Moreover, the AFM exchange interaction exhibits a high degree of sensitivity to biaxial strain, highlighting its crucial role in strain-mediated manipulation of magnetic properties. The SQUID-based magnetization measurements reveal the stabilization of the ferromagnetic ordering with compressive strain. The temperature-dependent evolution of the saturation magnetization field in the field-dependent magnetization measurement gives insights into the reorientation of spins towards a glassy phase. XAS and XMCD investigations at low temperatures confirm the mixed valency of Ni and Co ions, which are responsible for the ferromagnetic double exchange interactions. The element-specific magnetic moments extracted from the integrated XMCD spectra verify the AFM coupling between Pr 4\textit{f} and Co/Ni 3\textit{d} sublattices. The overall strain response to the magnetization dynamics was dominated by the Pr ordering, resulting in enhanced (reduced) AFM interactions in tensile (compressive) strained PCNO films. Further, the effect of strain tuning and Pr magnetism is also reflected in the field-dependent transport properties in PCNO films with stabilization of negative MR in PCNO/LAO and the temperature-dependent sign reversal of AMR in compressive strained films following the spin-reorientation-induced change in the magnetization easy axis. This work highlights the crucial role of rare-earth magnetism in controlling the spin dynamics of rare-earth cobaltite and nickelate oxide systems and how these complex magnetic ground states can be effectively tuned with strain. Furthermore, they demonstrate the effectiveness of strain engineering in tailoring complex magnetic ground states, offering a promising strategy for designing novel magnetic devices for spintronics applications.

\begin{acknowledgments}
M.S.R. and K.S. acknowledge the Science and Engineering Research Board (SERB) Grant
No. EMR/2017/002328, Department of Science and Technology, Government of India (DST-GoI) funding which led to the establishment of the Nano Functional Materials Technology Centre (NFMTC) [Grants No. SR/NM/NAT/02-2005 and No. DST/NM/JIIT-01/2016(C)], DST FIST-Phase II funding for PPMS [SR/FST/PSII-038/2016] and IIT Madras for establishing SVSM and Centre of Excellence (CoE) funding for QuCenDiEM [Grant No. SB20210813PHMHRD002720].

\end{acknowledgments}

\section{\label{appendix}Appendix}
\subsection{Normalized field dependent magnetic moment} 
\begin{table*}[t]
    \caption{Resistivity fit parameters for PCNO samples grown on STO and LAO substrates}
    \centering
    \begin{ruledtabular}
    \begin{tabular}{ccccccc}
    Substrate&\multicolumn{2}{c}{Arrhenius fit}&\multicolumn{2}{c}{3D Mott VRH fit}&\multicolumn{2}{c}{ES VRH fit} \\
    &$\rho_o (\Omega.cm)$&$E_a(meV)$&$\rho\textsuperscript{'}_o (\Omega.cm)$&$T_{Mott} (K)$&$\rho\textsuperscript{"}_o (\Omega.cm)$&$T_{ES} (K)$ \\ \hline\\
    (LaAlO$_3$)$_{0.3}$(Sr$_2$TaAlO$_6$)$_{0.7}$&$--$&$--$&$1.52\times10^{-5}$&$0.60\times10^6$&$3.35\times10^{-4}$&$3600$\\
    SrTiO$_3$&$--$&$--$&$1.26\times10^{-5}$&$1.09\times10^6$&$6.29\times10^{-4}$&$4186$\\
    LaAlO$_3$&$3.44\times10^{-3}$&$23.6$&$3.38\times10^{-8}$&$3.93\times10^6$&$--$&$--$\\ 
    \end{tabular}
    \end{ruledtabular}
    \label{tab:table2}
\end{table*}

\begin{figure}[ht!]
    \centering
    \includegraphics[width=\linewidth,keepaspectratio]{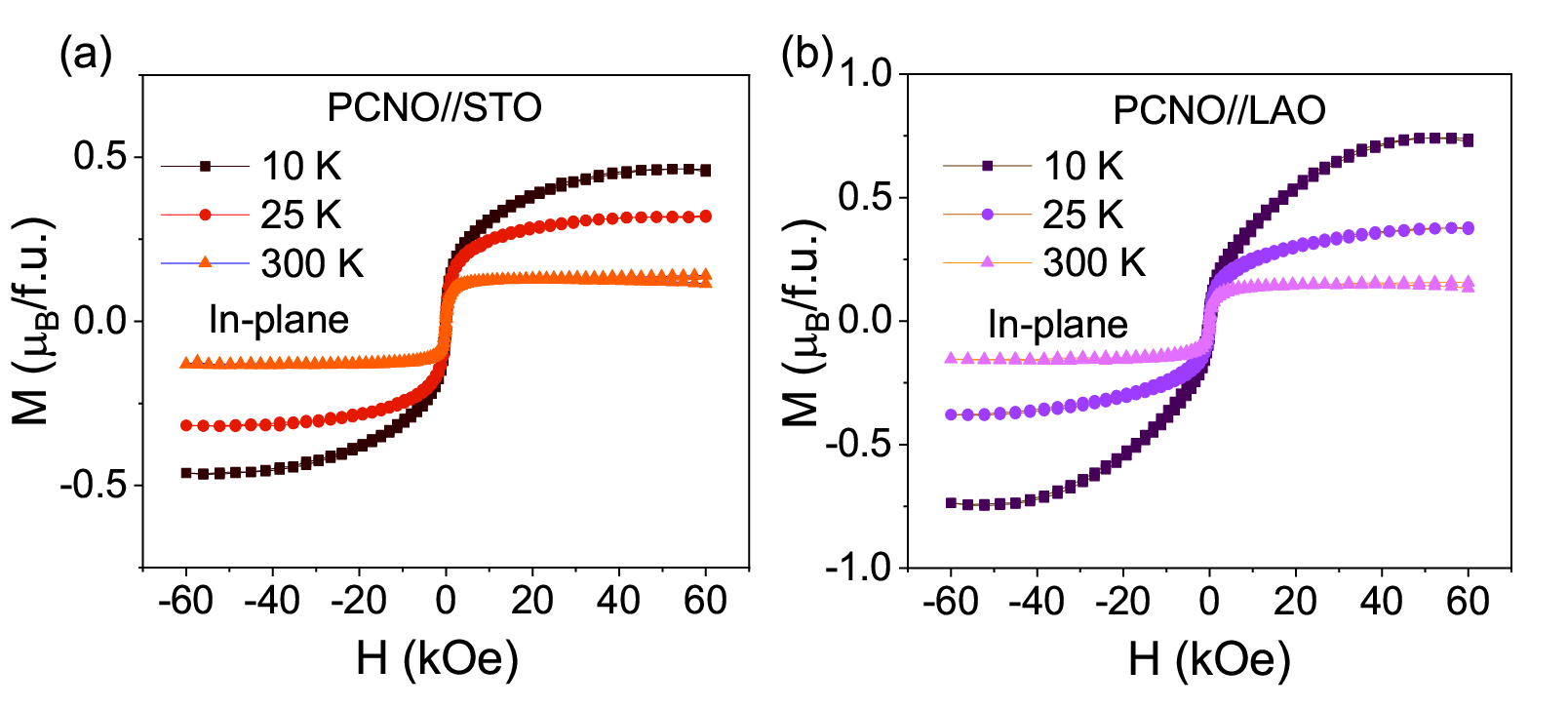}
\caption{\label{Fig:MH}Magnetization versus magnetic field curves for (a) PCNO/STO and (b) PCNO/ LAO films at different temperatures}
\end{figure}

Figure \ref{Fig:MH} gives the field dependent magnetization curves for PCNO/STO and PCNO/LAO films respectively. 

\subsection{Temperature dependence of electrical resistivity}

In solid solutions of PrNiO$_3$ and PrCoO$_3$, the temperature dependence of resistivity shows a semiconducting-like behaviour, where the low-temperature conduction is dominated by disorder-induced electron localization. The electrical transport in PrCo$_{1-x}$Ni$_x$O$_3$, reportedly follows the ES-VRH mechanism below 160 K, indicating strong electron localization effects \cite{tomevs2011transport}. 
In PCNO thin films, these localized electronic states are greatly tuned by biaxial strain, which affects the elemental bond angle and bond lengths within the perovskite layers.

\begin{figure}[ht]
    \centering
    \includegraphics[width=\linewidth,keepaspectratio]{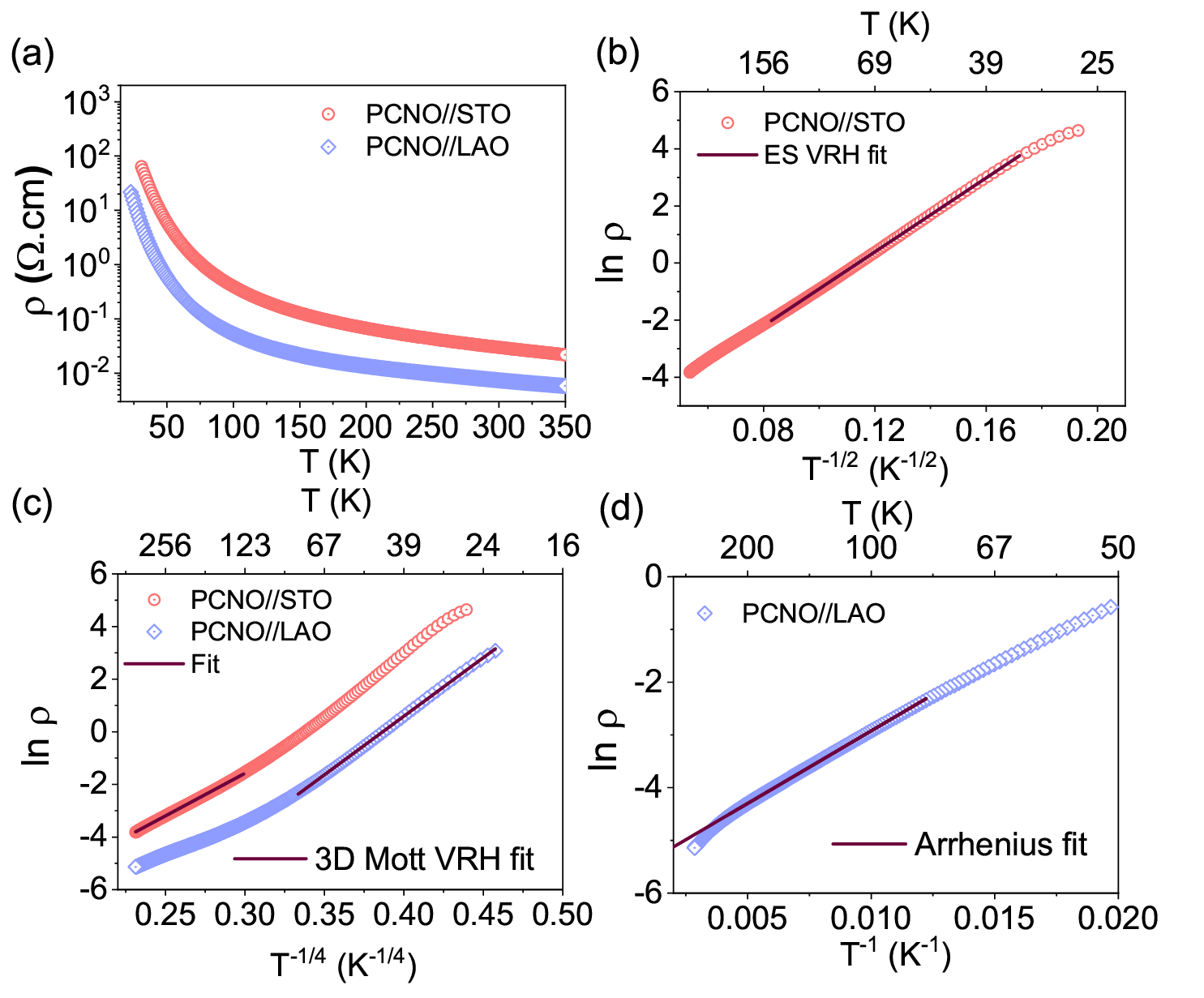}
\caption{\label{Fig:RT}(a) Semi-log scale plot of resistivity versus temperature for PCNO films grown on STO and LAO substrates. (b) Efros-Shklovskii variable range hopping fit for PCNO/STO at low temperatures, (c) 3D Mott variable range hopping fits for PCNO/STO and PCNO/LAO. (d) Arrhenius fit for the high-temperature resistivity data for PCNO/LAO film. Symbols represent the experimental data, and solid lines are the respective model fits}
\end{figure}

In order to understand the transport mechanism in strained PCNO thin films, temperature-dependent electrical resistivity measurements were done on films under tensile and compressive strain conditions. Figure \ref{Fig:RT} shows the temperature-dependent resistivity measurements for PCNO films subjected to tensile and compressive strain environments (i.e. PCNO/STO and PCNO/LAO, respectively). From Fig. \ref{Fig:RT} (a), it is clear that all films show semiconducting-like behaviour similar to bulk PCNO, with a steep increase in the magnitude of resistivity towards low temperatures. The order of magnitude of resistivity varies from $10^{-2}$ $\Omega$.cm at 300 K to as high as $10^2$ $\Omega$.cm close to 20 K. Such four-fold increase in the order of resistivity signifies increased localization and strong disorder-induced scattering. 

After considering various conduction mechanisms, the low-temperature resistivity data was best fitted with the 3D Mott variable range hopping (Mott-VRH) model as well as the Efros-Shklovskii variable range hopping model (ES-VRH) given by equations \ref{eqn:VRH} - \ref{eqn:TES} below, respectively. 

\begin{eqnarray}
\label{eqn:VRH}
\centering\rho(T)=\rho_o\exp
\left(
\frac{T_0}{T}
\right)^{1/n }
\end{eqnarray}
where n = 4 for 3D Mott VRH and 2 for ES VRH and $\rho_0$ takes their respective exponential prefactors of $\rho'_0$ and $\rho''_0$. Similarly the characteristic temperature, $T_o$ = $T_M$ for 3D Mott VRH, given as 
\begin{eqnarray}
\label{eqn:TMott}
\centering T_M = 
\frac{\beta_M}{k_B\xi^3D(E_F)}
\end{eqnarray}
and $T_o$ = $T_{ES}$ for ES VRH, given as 
\begin{eqnarray}
\label{eqn:TES}
\centering T_{ES} = 
\frac{\beta_{ES}e^2}{k_B\kappa\xi}
\end{eqnarray}

\noindent where we have adapted values of $\beta_M$ as 18.1 \cite{castner1991hopping} and $\beta_{ES}$ as 2.8 \cite{shklovskii1984correlation}. Here $D(E_F)$ is the density of states at the Fermi level, $\xi$ is the localization length and $\kappa$ is the dielectric constant.

For PCNO/LAO the high-temperature data was best fit with the Arrhenius model (\ref{eqn:Arrh}) given as 
\begin{eqnarray}
\label{eqn:Arrh}
\centering\rho(T)=\rho\textsuperscript{'}_o\exp
\left(
\frac{E_a}{k_BT}
\right)
\end{eqnarray}

\noindent where $E_a$ is the activation energy, $\rho_0$ is the exponential prefactor and $k_B$ is the Boltzmann constant.

Figure \ref{Fig:RT} (b)-(d) shows electrical resistivity fits with ES-VRH, 3D Mott-VRH and Arrhenius models, respectively. From the fit, it is clear that PCNO films follow variable range hopping at low temperatures, where conduction occurs through the hopping of electrons from one site to another energetically favourable site, which may not be its nearest neighbour. For large tensile strained film grown on STO (3.71\% strain), the conduction is completely dominated by the variable range hopping mechanism throughout the measured temperature range, with a crossover from 3D Mott-VRH to ES-VRH conduction at low temperature as shown in Fig.\ref{Fig:RT} (c) and (d). The crossover temperature is found using the relation\cite{RalphRosenbaum_1997}
\begin{eqnarray}
\label{eqn:T_VRH_cr}
\centering T_{cross}=\left(\frac{18.1 \pi T_{ES}^3}{3\times 2.8^3T_{M}}\right)^{1/2}
\end{eqnarray}

ES VRH to 3D Mott VRH crossover temperature in PCNO grown on STO is found to be 241 K.
The effect of biaxial compressive strain for films grown on LAO (-0.3\% compressive strain) tends to reduce the overall resistivity of PCNO relative to PCNO/STO film, and there is no crossover to ES VRH mechanism at low temperatures. Table \ref{tab:table2} summarizes the resistivity fit results for PCNO films on various substrates. It is found that the effect of strain has a considerable impact on electron hopping due to strain-induced changes in Co(Ni)-O bond lengths and Co(Ni)-O-Co(Ni) bond angles in the system. The origin of ES-VRH at low temperatures can be understood based on disorder-induced carrier localization, which is further enhanced by tensile strain-induced lattice stretching. For compressive strained PCNO films on LAO, the high-temperature resistivity fits well with the Arrhenius model given by equation \ref{eqn:Arrh}, indicating a thermally activated conduction for temperatures above 100 K and giving a band gap of 47.2 meV (see Fig. \ref{Fig:RT} (d) and Table \ref{tab:table2}. Below 100 K, the resistivity obeys the 3D Mott variable range hopping as the electrons get localized. Compressive strain tends to reduce electron localization in PCNO films, and the low temperature resistivity effectively follows the Mott-VRH instead of ES-VRH mechanism.


\bibliography{PCNO}

\end{document}